\documentclass[twocolumn,showpacs,preprintnumbers,amsmath,amssymb,superscriptaddress,prb]{revtex4}

\usepackage{graphicx}
\usepackage{bm}
\usepackage{mathenv}
\usepackage{color}
\usepackage{pifont}
\usepackage{setspace}
\usepackage{amsmath}

\begin{document}

\preprint{APS/123-QED}
\title{Revealing the ortho-II Band Folding in YBa$_{2}$Cu$_{3}$O$_{7-\delta}$ Films}
\author{Y.~Sassa}
\email{yasmine.sassa@psi.ch} \affiliation{Laboratory for Neutron Scattering, ETH Z\"{u}rich and Paul Scherrer Institut, CH-5232 Villigen PSI,
Switzerland}
\author{M.~Radovi\'{c}}
\affiliation{Laboratory for Synchrotron and Neutron Spectroscopy, EPFL, CH-1015 Lausanne, Switzerland} \affiliation{Swiss Light Source, Paul
Scherrer Institut, CH-5232 Villigen PSI, Switzerland}
\author{M.~M{\aa}nsson}
\affiliation{Laboratory for Neutron Scattering, ETH Z\"{u}rich and Paul Scherrer Institut, CH-5232 Villigen PSI, Switzerland}
\affiliation{Laboratory for Synchrotron and Neutron Spectroscopy, EPFL, CH-1015 Lausanne, Switzerland}
\affiliation{Laboratory for Solid state physics, ETH Z\"{u}rich, CH-8093 Z\"{u}rich, Switzerland}
\author{E.~Razzoli}
\affiliation{Laboratory for Synchrotron and Neutron Spectroscopy, EPFL, CH-1015 Lausanne, Switzerland} \affiliation{Swiss Light Source, Paul
Scherrer Institut, CH-5232 Villigen PSI, Switzerland}
\author{X.~Y.~Cui}
\altaffiliation[Present address: ]{University Duisburg-Essen Physics Department D-47048 Duisburg, Germany}
\affiliation{Swiss Light Source, Paul Scherrer Institut, CH-5232 Villigen PSI, Switzerland}
\author{S.~Pailh\`{e}s}
\affiliation{Laboratoire PMCN, UMR 5586, Universit\'{e} Lyon~1, CNRS-Universit\'{e} de Lyon, Villeurbanne 69622, France}
\author{S.~Guerrero}
\affiliation{Condensed Matter Theory Group, Paul Scherrer Institut, CH-5232 Villigen PSI, Switzerland}
\author{M.~Shi}
\affiliation{Swiss Light Source, Paul Scherrer Institut, CH-5232 Villigen PSI, Switzerland}
\author{P.~R.~Willmott}
\affiliation{Swiss Light Source, Paul Scherrer Institut, CH-5232 Villigen PSI, Switzerland}
\author{F.~Miletto~Granozio}
\affiliation{CNR-SPIN, Complesso Universitario Monte S. Angelo, 80126 Napoli, Italy}
\author{J.~Mesot}
\affiliation{Laboratory for Neutron Scattering, ETH Z\"{u}rich and Paul Scherrer Institut, CH-5232 Villigen PSI, Switzerland}
\affiliation{Laboratory for Synchrotron and Neutron Spectroscopy, EPFL, CH-1015 Lausanne, Switzerland}
\author{M.~R.~Norman}
\affiliation{Materials Science Division, Argonne National Laboratory, Argonne, IL 60439, USA}
\author{L.~Patthey}
\email{luc.patthey@psi.ch} \affiliation{Swiss Light Source, Paul Scherrer Institut, CH-5232 Villigen PSI, Switzerland}

\date{\today}

\begin{abstract}
We present an angle-resolved photoelectron spectroscopy study of YBa$_{2}$Cu$_{3}$O$_{7-\delta}$ films $in~situ$ grown by pulsed laser
deposition. We have successfully produced underdoped surfaces with ordered oxygen vacancies within the CuO chains resulting in a clear ortho-II
band folding of the Fermi surface. This indicates that order within the CuO chains affects the electronic properties of the CuO$_{2}$ planes.
Our results highlight the importance of having not only the correct surface carrier concentration, but also a very well ordered and clean
surface in order that photoemission data on this compound be representative of the bulk.
\end{abstract}

\maketitle
Since the discovery of high-temperature superconductors (HTSC), the YBa$_{2}$Cu$_{3}$O$_{7-\delta}$ (Y123) compound has been the
subject of many experimental and theoretical studies \cite{Hossain, Doiron, Carrington, Elfimov}. Recently, interest in this material was
renewed when quantum oscillation experiments in high magnetic fields in underdoped ortho-II ordered YBa$_{2}$Cu$_{3}$O$_{6.5}$ (YBCO$_{6.5}$)
revealed that the Fermi surface (FS) reconstructs into one \cite{Doiron} or several \cite{Audouard} pockets. Indeed, theoretical predictions
have shown that ortho-II ordered YBCO$_{6.5}$ should display band folding giving rise to pockets \cite{Carrington, Elfimov}. Further, for other
cuprate HTSC, evidence for such pockets have been reported by angle-resolved photoelectron spectroscopy (ARPES) \cite{Chang, Razzoli}. However,
for ordered YBCO$_{6.5}$, no pockets or band folding have been directly observed by ARPES \cite{Hossain}.

The crystal structure of Y123 [Fig.~1(a)] differs slightly from other HTSC. In addition to the CuO$_{2}$ planes, it also contains 1D CuO chains
along the $b$-axis that donate charge carriers (holes) to the superconducting planes. It is also well established that Y123 displays a wide
variety of superstructures \cite{Wang} caused by oxygen-vacancy order within the chains. One in particular is the ortho-II phase, characterized
by ordered oxygen-vacancies within every second CuO chain [Fig.~1(b)]. This alternation of filled and empty chains along the $b$-axis induces a
unit-cell doubling along the $a$-axis, i.e., a reduction of the Brillouin zone, and hence band folding is expected \cite{Carrington, Elfimov}.
Unfortunately, ARPES experiments on Y123 and especially the ortho-II phase are notoriously difficult. One reason is that the crystal structure
lacks a natural cleavage plane and hence cleaved, the surface contains both CuO and BaO terminations, giving different contributions to the
total ARPES intensity \cite{Zabolotnyy1}. Moreover, due to polarity, the cleaved surface tends to be strongly overdoped
\cite{Zabolotnyy1, Hossain} even though the bulk is underdoped. As a result, details of the electronic properties have remained elusive. To
avoid the problem with self-hole-doping of Y123, Hossain $\textit{et al.}$ \cite{Hossain} performed an $in~situ$ evaporation of potassium (K)
onto the cleaved surface. Although the correct hole-doping was achieved, the importance of oxygen-vacancy ordering as well as surface termination
remained unresolved and the expected ortho-II band folding \cite{Carrington, Elfimov} was not detected. Due to the absence of experimental
evidence, the ortho-II potential was assumed to be too weak, and such band folding was not considered in many theoretical models
\cite{Harrison}.
\begin{figure}
\includegraphics[keepaspectratio=true,width=70 mm]{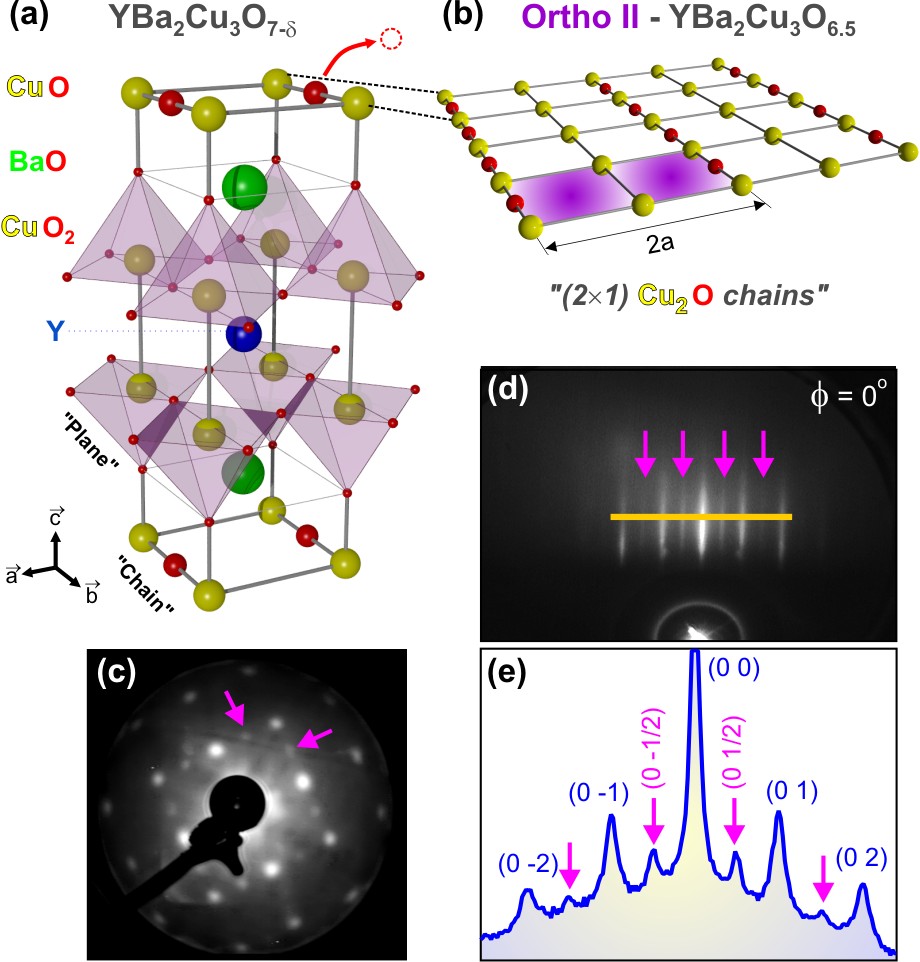}
\caption{\label{fig:1}(color online) (a) Structure of Y123 showing the CuO$_{2}$ planes and CuO chains. (b) Half-filled and ordered (2$\times$1)
ortho-II chains corresponding to $n$~=~6.5. (c) Low-energy electron diffraction (LEED) pattern shows extra spots from the (2$\times$1) chain
order (arrows) in both directions because of twinning. (d) Reflection high-energy electron diffraction (RHEED) pattern. (e) Cut along the solid
line in (d). Superstructure peaks at e.g. (0 -1/2) and (0 1/2) (arrows) indicate the unit-cell doubling.}
\end{figure}

In this rapid communication we present a high-resolution ARPES study of heteroepitaxial Y123 films, grown $in~situ$ by pulsed laser deposition
(PLD) \cite{Willmott}. The ARPES experiments were performed on the Surface/Interface Spectroscopy (SIS) X09LA beamline at the Swiss Light
Source, Paul Scherrer Institut, Villigen, Switzerland. The beamline was set to circular polarized light with a photon energy, $h\nu$~=~70~eV and
data were acquired using both Gammadata Scienta SES-2002 and VG-Scienta R4000 electron analyzer. The energy resolution was set to 15-25~meV
and the momentum resolution parallel/perpendicular to the analyzer slit chosen as $\sim$0.009/0.019~\AA$^{-1}$. Data were acquired in the
temperature range $T$~=~9-120~K using the six degree of freedom CARVING manipulator. The binding-energy scale was calibrated with a copper
reference sample in direct electrical and thermal contact with the film. The base pressure of the UHV system was below 5$\times$10$^{-11}$~mbar
during the entire measurement and no sign of sample/data quality degradation was observed. Our results were reproduced on several occasions,
using more than ten different samples grown under the same conditions.

The 100~nm thick Y123 films were grown on TiO$_{2}$ (B-site) terminated SrTiO$_{3}$ (STO) substrates, resulting in a CuO chain termination of
the Y123 surface \cite{Huijbregtse}. Reflection high-energy electron-diffraction (RHEED) measurements of the films show a streaky pattern
[Fig.~1(d)], suggesting a clear 2D growth with an atomically flat crystalline surface. Moreover, additional peaks around e.g. (0~1/2) and
(0~-1/2) [Fig.~1(e)] can be distinguished, suggesting a unit-cell doubling. Also the low-energy electron diffraction (LEED) pattern [Fig.~1(c)]
displays a clear (2$\times$1) reconstruction, as indicated by the extra weaker spots. It is also clear that the film is twinned since the
(2$\times$1) spots appear in both directions. The bulk properties of the film were verified $ex~situ$ by X-ray diffraction (XRD) measurements.
The XRD pattern shows a very good $c$-axis orientation and a narrow (FWHM~$\approx$~0.08$^{\circ}$) rocking curve of the YBCO (005) reflection
(not shown).

\begin{figure}
\includegraphics[keepaspectratio=true,width=85 mm]{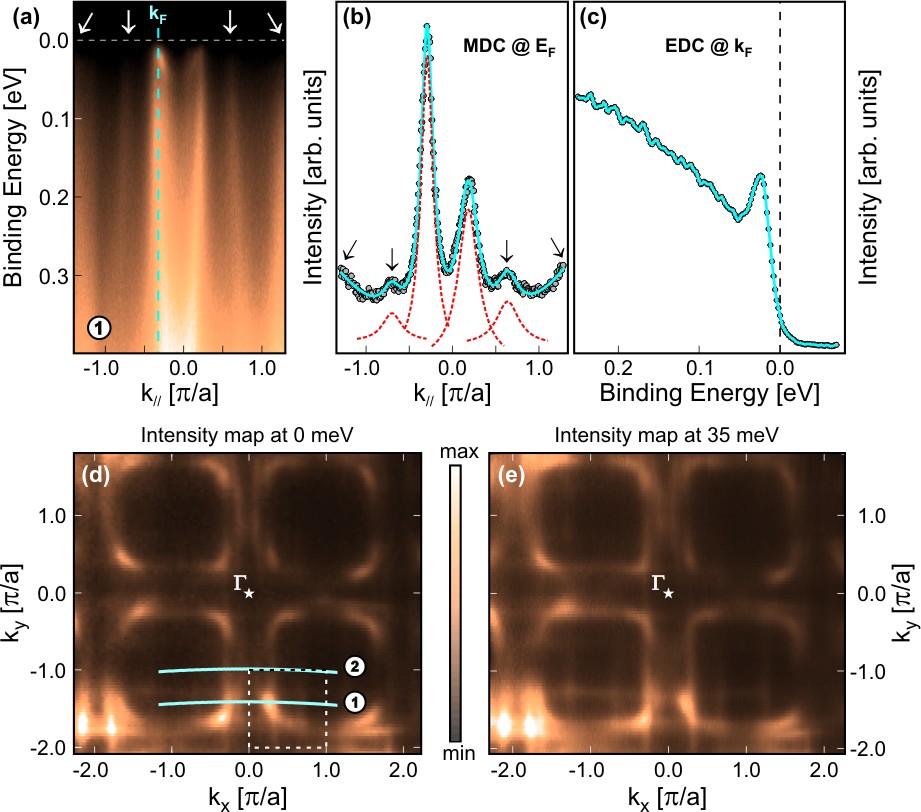}
\caption{\label{fig:2}(color online) (a-b) ARPES spectra acquired at cut \ding{172} in $k$-space as indicated by the solid line in (d). (b)
Momentum distribution curve (MDC) at the Fermi level ($E_{\rm F}$) and a Lorentzian fit (solid and dashed lines). (c) Energy distribution curve
(EDC) at $k_{\rm F}$ and $T$~=~10~K. (d) Spectral intensity map obtained by energy integration of ARPES spectra $\pm$~5~meV about $E_{\rm F}$.
The two solid white lines mark two specific cuts \ding{172} \& \ding{173} in $k$-space. (e) Same as (d) but $\pm$~5~meV about $E_{\rm
B}$~=~35~meV. From photon-energy dependent measurements, a photon energy of $h\nu$~=~70~eV was chosen in order to maximize the
spectral intensity from the CuO$_{2}$ planes while suppressing the contribution from the CuO chains.}
\end{figure}

By a direct $in~situ$ transfer between the PLD and ARPES UHV chambers, we have been able to measure the Y123 films \emph{as grown} (without
cleaving). In Fig.~2(a) a typical ARPES spectrum is shown, that was acquired at cut \ding{172} [Fig.~2(d)] and $T$~=~10~K. Two strong dispersive
features are clearly observable but also four supplementary weaker bands are visible, that were not present in previous ARPES measurements on
Y123 \cite{Hossain}. From the dispersion it is clear that the supplementary bands are created by a folding of the main bands. This is consistent
with the extra lines/points in the RHEED/LEED patterns (Fig.~1) that reveal a unit-cell doubling. The folded bands are even more visible in
Fig.~2(b) where the result of a Lorentzian fit of the momentum distribution curve (MDC) at the Fermi level ($E_{\rm F}$) is shown. Fig.~2(c)
represents the energy distribution curve (EDC) at the Fermi wavevector ($k_{\rm F}$) for one of the main bands in Fig.~2(a) (blue dashed
vertical line). A sharp quasi-particle (QP) peak is visible slightly below $E_{\rm F}$, indicating the presence of a superconducting gap.

By acquiring ARPES spectra for multiple momentum cuts, we have mapped out the FS of the Y123 film. Fig.~2(d) shows the spectral intensity map
integrated $\pm$~5~meV around $E_{\rm F}$ and Fig.~2(e) the underlying FS integrated $\pm$~5~meV around a binding energy $E_{\rm B}$~=~35~meV
(i.e., below the gap). Centered around ($\pi$, $\pi$), we find a weaker pocket-like feature (square) associated with the folded bands and
unit-cell doubling, as mentioned above. By extracting the surface-hole doping ($p_{\rm surf}$) from the FS area $p_{\rm surf}$~=~0.1~$\pm$~0.02
is obtained, corresponding to an oxygen content ${n}$~=~7~-~$\delta{}\approx$~6.5, i.e. YBCO$_{6.5}$. It can be concluded that the surface of
the film has both the correct hole doping and the necessary (vacancy) order to be a good representation of ortho-II YBCO$_{6.5}$.

Fig.~3(a) represents the EDC at $k_{\rm F}$ (blue dashed line showed in Fig.~2(a)) as a function of temperature. At low temperature
(10~K$\leq{}T\leq$60~K), a clear QP peak is observed. At higher $T$, this peak decreases significantly and almost vanishes at $T$~=~70~K. The
appearance of such a coherent peak below $T_{\rm c}$ is a good indication for the presence of a superconducting state. In order to estimate the
superconducting gap-value, the EDCs are symmetrized with respect to $E_{\rm F}$ [Fig.~3(b)]. The extracted gap size for cuts \ding{172} at
$T$~=~10~K is $\Delta_{1}$~=~22~$\pm$~3~meV. Increasing $T$ reduces the gap value until it closes at $T$~=~70~K, indicating that $T_{\rm
c}\approx$~65~K. From the value of $T_{\rm c}$ it is deduced that $p^{T_{\rm c}}$~$\approx$~0.106 \cite{Liang}, which is in good agreement with
the value extracted from the FS area. Fig.~3(c) shows the EDC at $k_{\rm F}$ at the antinode for cut \ding{173} [Fig.~2(d)] at $T$~=~10-120~K.
Contrary to cut \ding{172}, the QP peak at $T$~=~10~K is strongly suppressed, which is a common feature of underdoped cuprates. Like Fig.~3(b),
the EDCs are symmetrized and the obtained gap size is $\Delta_{2}$~=~52~$\pm$~4~meV. The values of $\Delta_{1}$ and $\Delta_{2}$ are
consistent with a strongly anisotropic gap having its maximum at the antinodal point. Above the superconducting transition at $T$~=~120~K, the
symmetrized EDCs show that $\Delta_{2}$ remains open, while $\Delta_{1}$ has closed at $T$~=~70~K. This strongly suggests the existence of a
pseudogap state, again confirming that the surface of the film is indeed underdoped.

\begin{figure}
\includegraphics[keepaspectratio=true,width=73 mm]{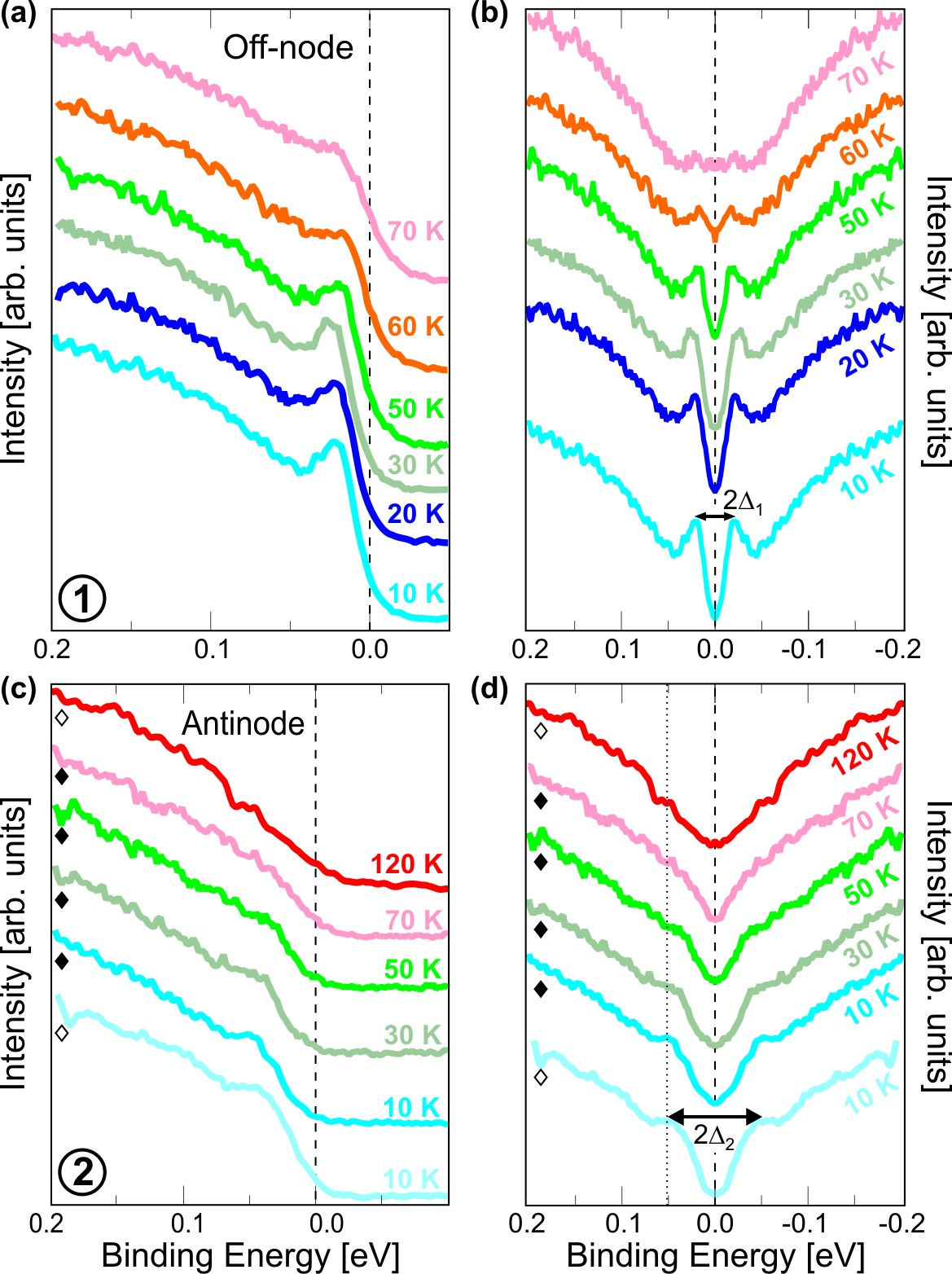}
\caption{\label{fig:3}(color online) (a) Energy distribution curves (EDCs) for cut \ding{172} in Fig.~2(d) as a function of temperature,
$T$~=~10-70~K. (b) Symmetrization of the EDCs in (a) showing the opening of the superconducting gap ($\Delta_{1}$~=~22~$\pm$~3~meV) below
$T_{\rm c}\approx$~65~K. (c) EDCs for cut \ding{173} in Fig.~2(d) for $T$~=~10-120~K. $\Diamond$ and \ding{169} are the results from two
different samples. (d) Symmetrization of the EDCs in (c) showing the presence of a superconducting gap below $T_{\rm c}$ and a pseudogap above
$T_{\rm c}$ ($\Delta_{2}$~=~52~$\pm$~4~meV). }
\end{figure}

To consider the effect of the ortho-II order on the electronic structure as only a simple band folding along the $a$-axis is a \emph{highly}
simplified picture. For instance, it has been shown by nuclear magnetic resonance (NMR) that the vacancy order causes a charge imbalance between
the Cu atoms sitting below filled/empty chains \cite{Yamani}. Naturally, a coupling between the bands of the chains and the superconducting
CuO$_{2}$ planes is expected. Bascones $et~al.$ theoretically demonstrated that for ortho-II, due to interlayer coherence and oxygen ordering,
the bonding (B) and anti-bonding (AB) bands of Y123 are each split into two bands, $\alpha$ and a quasi-one-dimensional $\beta$ due to zone folding
\cite{Bascones}. Taking into account the 2$a$ periodicity of the ortho-II phase, the resulting dispersion of the CuO$_{2}$ plane can be expressed
according to \cite{Bascones}:
\begin{eqnarray}
& & \varepsilon^{\rm{AB,B}}_{\alpha,\beta}(\rm \textbf{k}) =
-2\emph{t}~\cos{\emph{k}_{\emph{y}}}-2\emph{t''}\left(\cos2\emph{k}_{\emph{x}}+\cos2\emph{k}_{\emph{y}}\right) \cr &&-\mu{\rm
\pm}~t_{\perp}\textbf{\rm (\textbf{k})}{\rm \pm}~\left[4 \cos^{\rm 2}k_{x}(t-2t' \cos{}k_{y})^{\rm 2}+\frac{V^{\rm 2}}{4}\right]^{\rm
\frac{1}{2}}\cr \label{eq:TB}
\end{eqnarray}
where $\mu$ is the chemical potential, $t$ is the nearest neighbor hopping integral, $t'$ and $t''$ the second and third nearest-neighbor
intraplane hopping integrals, $t_{\rm \perp}$ the interlayer hopping integral (bilayer splitting), and $V$ the ortho-II potential. The first and
second $\pm$ signs set the AB/B bands and the $\alpha$/$\beta$ bands, respectively. Here, $V$ is set constant since from local density
approximation (LDA) calculations, the ortho-II potential is only slightly $k$-dependent near the nodes \cite{Bascones}. Further, since the
existence of bilayer splitting in the underdoped regime remains unclear and cannot be distinguished in our data, $t_{\rm \perp}$ is neglected.
This is consistent with very recent ARPES data recorded for different dopings of Y123 surfaces, which demonstrated that the bilayer splitting is
progressively reduced upon underdoping and vanishes below $p$~$\approx$~0.15 \cite{Fournier}. The FS can be adequately fitted by the
tight-binding model described by Eq.~(\ref{eq:TB}), giving $t$~=~558~$\pm$~50~meV, $t'/t$~=~0.49~$\pm$~0.03, $t''/t'$~=~0.5~$\pm$~0.03,
$\mu$~=~-469~$\pm$~90~meV and $V$~=~75~meV. Fig.~4(a) shows the calculated FS of the ortho-II sample considering the tight-binding parameters
mentioned above. It is clear that the reduction of the Brillouin zone induces a drastic change in the shape but also the number of FS sheets.
For a twinned sample, the FS will be folded along both the ($\pi/2,0$)-($\pi/2,\pi$) as well as the ($0,\pi/2$)-($\pi,\pi/2$) lines. Fig.~4(b)
shows the calculated FS for a twinned ortho-II sample and Fig.~4(c) the corresponding simulated intensity map at 0~meV with an energy broadening
of 25~meV. In the latter, the upper-right quadrant represents the experimental data [dashed square area in Fig.~2(d)] and the filled circles are
extracted from Lorenztian fits of the MDCs at $E_{\rm F}$ for different $k$-momenta. Fig.~4(d) shows the filled circles (symmetrized in momentum)
along with the calculated twinned FS. By comparing the calculated intensity map with the data [Fig.~4(c)] and by overlaying the model onto the
experimental points [Fig.~4(d)] it is clear that there is very good agreement.

\begin{figure}
\includegraphics[keepaspectratio=true,width=84 mm]{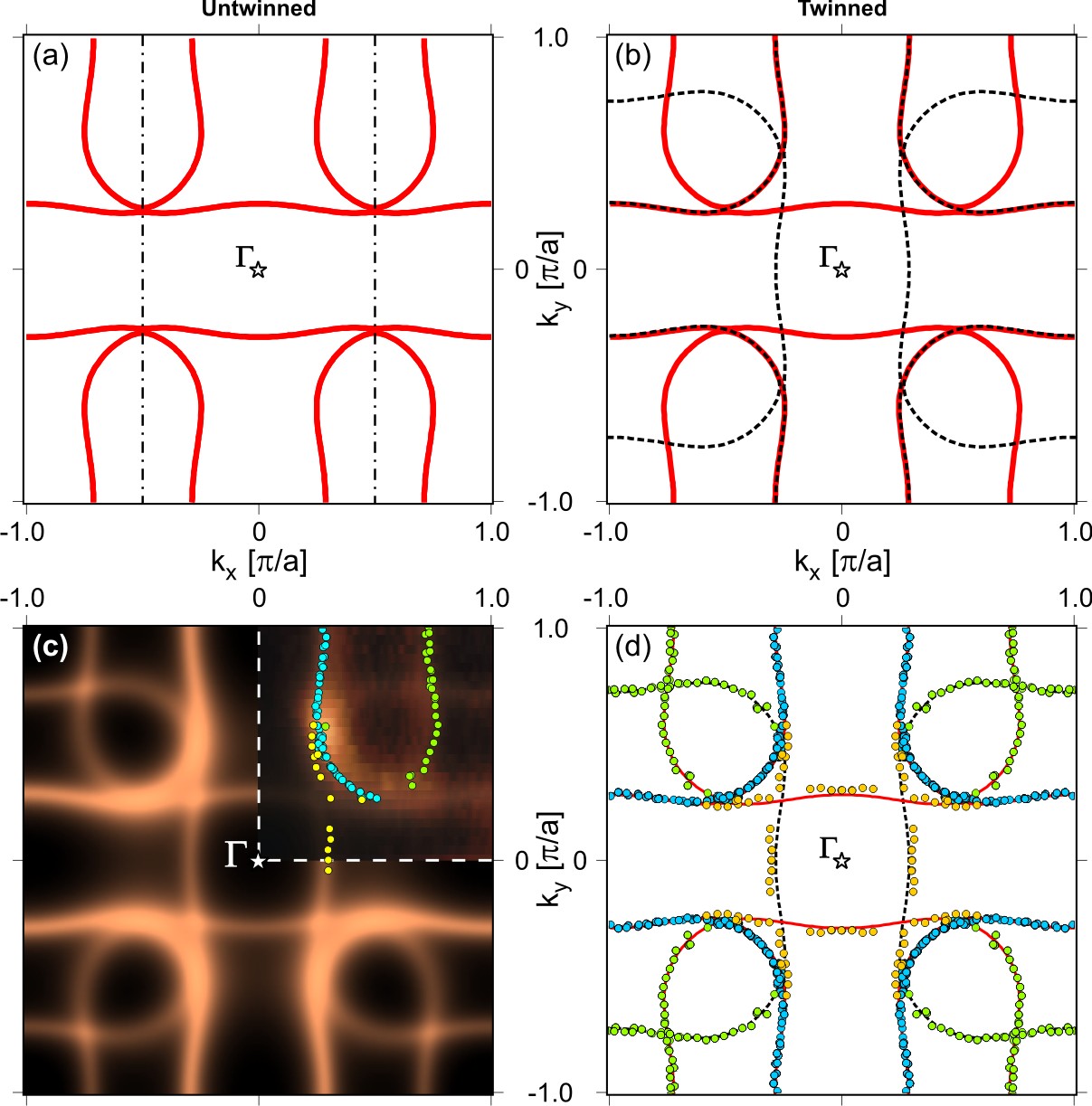}
\caption{\label{fig:4}(color online) (a) Tight-binding Fermi surface (FS) for ortho-II YBCO$_{6.5}$ calculated using Eq.~(\ref{eq:TB}). (b) Same
as (a) but after the addition of a second, 90$^{\circ}$ rotated, domain caused by twinning (dashed lines). (c) Simulated spectral intensity map
at $E_{\rm F}$. The upper right quadrant shows the experimental data [extracted from the dashed square in Fig.~2(d)] and the filled circles represent $k_{\rm F}$ as determined from Lorenztian fits of the MDCs at $E_{\rm F}$. (d) Overlay of experimentally determined $k_{\rm F}$ (filled circles) and calculated ortho-II FS from (b).}
\end{figure}

In summary, we have been able to measure Y123 films by ARPES, a task previously thought not possible due to oxygen deficiency causing an
insulating surface \cite{Abrecht}. By growing oxygen-ordered Y123 films $in~situ$, a clear surface representation of ortho-II band folding is
made evident by ARPES. Our results thereby confirm theoretical \cite{Carrington, Elfimov} and experimental expectations \cite{Feng}. We connect this
to a (2$\times$1) surface reconstruction caused by ordered oxygen vacancies that help to stabilize the Y123 surface. Our experiments clearly
highlight the importance of having not only the correct carrier concentration, but also a very well ordered and clean surface to facilitate
ARPES data representative of the compound's true nature. This could also explain why the ortho-II band folding was not found by Hossain $et~al.$
\cite{Hossain}. Even though they evidently managed to obtain the correct (under)doping of the surface by K evaporation, the essential ordering
of oxygen-vacancies was not fulfilled. It has for other compounds been shown that charge carrier concentration $and$ ion-vacancy ordering are of
equal importance for the magnetic \cite{Schulze} as well as electronic properties \cite{Balicas}. In this case, we have successfully shown how oxygen vacancy
ordering in the CuO chains of the surface influences the electronic properties of the superconducting CuO$_{2}$ planes in Y123 films. In fact,
such unidirectional band folding could contribute to the strong breaking of the four-fold rotational symmetry of the CuO$_{2}$ planes observed
by the Nernst effect \cite{Daou}. This work opens the door to directly investigate such matters by performing ARPES measurements on, preferably,
untwinned Y123 samples. We would also like to emphasize that having obtained a more accurate view of the \emph{bulk} electronic structure in
underdoped Y123, the ortho-II folding can no longer be ignored \cite{Hossain, Harrison}. Instead our results give solid experimental support for
theoretical models that consider a combination of ortho-II band folding and magnetic \cite{Oh, Carter} or $d$-density wave (DDW) order
\cite{Podolsky}, which could be important for explaining the quantum oscillation results \cite{Doiron, Audouard}.
\\
We are grateful to Andrea Damascelli for valuable discussions, as well as Christian M. Schlep\"{u}tz for his support. This research was
supported by the Swiss National Science Foundation, MaNEP, and the Foundation BLANCEFLOR Boncompagni-Ludovisi n\'{e}e Bildt. MN was supported by
the US DOE, Office of Science, under contract DE-AC02-06CH11357.



\end{document}